\documentclass{article}

\usepackage[british]{babel}
\usepackage[useregional]{datetime2}
\DTMlangsetup[en-GB]{showdayofmonth=false}
\usepackage[numbers, sort]{natbib}
\usepackage[a4paper,top=3cm,bottom=3cm,left=3cm,right=3cm,marginparwidth=1.75cm]{geometry}
\usepackage{amsmath}
\usepackage{amsfonts}
\usepackage{amssymb}
\usepackage{mathtools}
\usepackage{wrapfig}
\usepackage{graphicx}
\usepackage{tikz}
\usepackage[colorlinks=true, allcolors=blue]{hyperref}
\usepackage[page]{appendix} 

\usepackage{graphicx}
\usepackage{subcaption}
\usepackage{multicol}
\usepackage{lipsum}

\renewcommand{\vec}[1]{\boldsymbol{\mathbf{#1}}}
\renewcommand{\v}{\vec}

\newcommand{\R}[1]{{R_{#1}}}
\newcommand{\w}[1]{w_{#1}}
\newcommand{\D}[1]{\Delta_{#1}}
\newcommand{\dw}[1]{\Delta w_{#1}}

\makeatletter
\newif\ifnobrackets
\renewcommand\@cite[2]{\ifnobrackets\else[\fi{#1\if@tempswa , #2\fi}\ifnobrackets\else]\fi\nobracketsfalse}

\makeatother
\usepackage{authblk}

\title{\vspace*{0cm}{\textbf{\Large{Multiblock MEV opportunities \& protections in dynamic AMMs}}}}
\author{Matthew Willetts}
\author{Christian Harrington}
\affil{QuantAMM.fi}
\begin{document}
\maketitle

\begin{abstract}
Maximal Extractable Value (MEV) in Constant Function Market Making is fairly well understood.
Does having dynamic  weights, as found in liquidity boostrap pools (LBPs), Temporal-function market makers (TFMMs), and Replicating market makers (RMMs), introduce new attack vectors?
In this paper we explore how inter-block weight changes can be analogous to trades, and can potentially lead to a multi-block MEV attack.
New inter-block protections required to guard against this new attack vector are analysed.
We also carry our a raft of numerical simulations, more than 450 million potential attack scenarios, showing both successful attacks and successful defense.

\end{abstract}
\section{Introduction}
\label{ssec:outline_attack}

In \emph{Temporal-Function Market Making} (TFMM)~\cite{tfmm_litepaper} the trading function of a pool depends on time, in contrast with Constant-Function Market Making.
In a TFMM pool the allocation of assets is responsive to market information.
The changing trading function of TFMM pools creates arbitrage opportunities (via changes in pools's quoted prices) that incentivises pools' holdings to be rebalanced by external agents to the pools' target holdings.

An important question about TFMM pools is whether their (known) temporal dependence leads to an opportunity for maximal extractable value (MEV).\footnote{a.k.a. Miner Extractable Value}
Here we study such attacks, where an attacker aims to manipulate a TFMM pool shortly before a weight update.
This situation is the TFMM equivalent of a `sandwich attack', but with the attack being done against pool LPs rather than a trader.

It is important to note that changes in trading function happen \emph{between blocks}.\footnote{This makes the weight updates somewhat reminiscent of trades in \href{https://www.paradigm.xyz/2021/07/twamm}{Time-weighted average market makers}.}
This means that any attack would necessarily be multi-block.
We perform mathematical analysis of the circumstances where attacks of this kind are possible and where they are not possible, even when granting the attacker (for free and with certainty) the final transaction in one block and the first transaction in the next.




This understanding motivates three extremely simple `guardrails' for protection against such manipulation:
\begin{itemize}
    \item limit on the size of weight changes,
    \item limit of trades as a fraction of pool reserves,
    \item limit on minimum allowed value in weights $\v w(t)$.
\end{itemize}
We have to make certain assumptions to obtain our analytical results.
Making use of recent advances in the modelling of optimal multi-token arbitrage trades in G3Ms~\cite{willetts2024closedform}, we carry out a battery of numerical simulations of potential attacks ($\sim 450,000,000$) to test the effectiveness of the proposed guardrails in rendering the potential multi-block attack unviable.


\section{Background: Temporal Function Market Making} \label{sec:tfmm-trading}
\emph{Temporal-Function Market Making} (TFMM)~\cite{tfmm_litepaper} pools have time-varying trading functions.
This is naturally achieved by having $\v w$, the portfolio vector of a G3M pool~\cite{balancer} change from block to block.
(For a pool of $N$ assets, each with reserves $R_i$, the G3M trading function is $\prod_{i=1}^N R_i^{w_i}= k$ where $k$ is the pool's constant.)
From the temporal dependency of $\v w$ in TFMM pools, we now have $\v w(t)$, and the (now time-varying) trading function is:
\begin{equation}
    \prod_{i=1}^N R_i^{w_i(t)}= k(t), \quad\mathrm{where\,} \sum_{i=1}^N w_i(t) = 1, \,\,\mathrm{and\,\,} \forall i\,\, 0< w_i(t)<1.
    \label{eq:tfmm}
\end{equation}
where we have made the time-dependence of $k$ explicit, as it now depends on the current, changing, value of $\v w(t)$.
Trades must preserve (or increase) the value of $k(t)$ that exists at the time of that trade.\footnote{Note that changes in weights from block to block do change the value of $k(t)$. This makes it hard to interpret a TFMM pool's $k(t)$ as an `unscaled pool value', as $k$ can be in vanilla G3Ms.}
Within a block weights are constant (i.e. $t$ is the discrete block number) and so take a known, clear, value at the time that any trade happens.


A TFMM pool thus takes a dynamic position, re-weighting its desired division of value between assets on the fly, doing so via offering arbitrage opportunities if its desired reserves are not in line with its actual reserves.


As described in~\cite{tfmm_litepaper}, multi-token trades are allowed on TFMMs.
Consider a trader wishing to exchange a particular set of tokens, represented by the vector $\v \Delta$, for another set of tokens $\v \Lambda$, where entries in each are $\geq 0$.
The pool requires, for tokens $i$ where $\Delta_i>0$, that $\Lambda_i=0$: tokens cannot be traded for themselves.
With fees $(1-\gamma)$, for a trade to be accepted it must be the case that
\begin{equation}
    \prod_{i=1}^N \left(R_i + \gamma \Delta_i - \Lambda_i\right)^{w_i(t)} \geq k(t)
    \label{eq:tfmm_w_fees}
\end{equation}

\paragraph{Token-Pair Quotes via the Trading Function}
Consider token-pair trades, i.e. $\v \Delta$ and $\v \Lambda$ are restricted to be one-hot.
The trader wishes to exchange $\Delta_i$ of the $i^\mathrm{th}$ token for some amount $\Lambda_j$ of the $j^\mathrm{th}$ token, $i\neq j$.
We can directly give a {TFMM} pool's quote for the amount $\Lambda_j$ that it will accept for the trading-in of $\Delta_i$, via inversion of Eq~\eqref{eq:tfmm_w_fees}:
\begin{equation}
    \Lambda_j = R_j \left(1 - \frac{1}{\left(1+\gamma \frac{\Delta_i}{R_i}\right)}^{w_i(t)/w_j(t)} \right).
    \label{eq:pair_trade}
\end{equation}
From taking a binomial expansion of Eq~\eqref{eq:pair_trade} for small $\Delta_i$ and setting $\gamma=1$ we can find the effective price quoted by a TFMM pool for the $i^\mathrm{th}$ token in terms of the $j^\mathrm{th}$ (that is, using the $j^\mathrm{th}$ token as the num\'eraire) if no fees are charged,
\begin{equation}
    p^{\mathrm{TFMM}}_{i,j}(t) = \frac{\frac{w_i(t)}{R_i}}{\frac{w_j(t)}{R_j}}.
    \label{eq:prices_numeraire}
\end{equation}
This result gives relationship between a pool's weights and the prices it quotes, and is of the same form as for G3Ms but for the time-dependence of $\v w$.
If the weight of token $i$ increases relative to token $j$, the price of token $i$ increases relative to token $j$; conversely, a decrease in the relative weight of token $i$ causes a decrease in its price, such that trading reduces holdings of token $i$ and increases those of $j$.
\newpage
\section{Multiblock MEV and TFMM weight changes}

The weights of a pool change from block to block.
This means that, even if the reserves of the pool are unchanged, the quoted prices of the pool change from block to block.

Knowing that the upcoming change in pool weights, and thus quoted prices, will lead a pool to offering an above-market price when buying one of its constituents, an attacker could, prior to the weight change, trade with the pool to reduce its holdings of that asset.
That trade would increase the quoted price of that asset, which would then stack with the weight-change-induced increase in quoted price.
The attacker could then, in the next block, sell this asset, at this now higher price, back to the pool.
In effect, by trading with the pool prior to the weight change, to manipulate the pool's quoted prices, the attacker has supercharged the arbitrage opportunity caused by the weight change and thus extract value from the pool.

This attack is intrinsically multi-block, and if the attacker does not correctly obtain the last transaction in the initial block and the first transaction in the second block they themselves are liable to be frontrun or sandwiched.
In our analysis here, we give the attacker costless and certain placement of their transactions in the way needed for the attack.

This potential attack is a three-step process.
To analyse it, we need to study the cost of manipulating the quoted prices of a pool, the weight change itself, and the return from performing an arbitrage trade against the pool's new weights.
From these quantities we can then calculate the overall return on this potential attack.

First, we derive the cost of manipulating a TFMM pool, with particular weights, such that it quotes a chosen price.
We do this in the presence of fees.
After having manipulated the price in this way, the weights of the pool change, and a `boosted' arbitrage trade is then available in the following block.
We derive the returns to an arbitrageur from bringing the manipulated-pool back to quoting market prices after the weights have been updated.

Later in paper we describe a battery of in-silico experiments where we perform this attack on TFMM pools, optimising not only the attacking trades but the state (weights, weight changes, reserves) of the pools themselves so as to make the attack as good as possible within the confines of the pools' `guardrail parameters' listed above.
For various settings of these guardrail parameters, attacks are not obtained for any settings of pool state when optimising the pool state (within the confines of the guardrails) for an attack to be found.







\subsection{Trading \& Fees}
\label{ssec:trading_and_fees}
Consider an $N$-token pool.
For simplicity of analysis we trade a subset of 2 tokens for this attack---in our numerical work later we allow all tokens to be traded and empirically we find that optimal attacks are done by trading between a pair of tokens in this way.
Those two tokens have weights $\v w=\{w_1,w_2\}$ and reserves $\v R=\{R_1,R_2\}$.
We are here going to consider the quoted prices of the tokens in the pool.

We assume the existence of true, external market prices for the tokens.
$m_{p,2}=m_p$ is the market price of the second token in terms of the first.
$m_{p,1}$, the market price of the first token in terms of the second is simply the reciprocal of this: $m_{p,1}=1/m_{p,2}=1/m_p$.
(Without loss of generality, we will be using the first token as the num\'eraire throughout unless otherwise noted.\footnote{Either you can imagine that the first token in the pool is a stablecoin in USD, say; or it can be an arbitrary token, and we then are applying a linear scaling (that token's price) to our vanilla market prices to get them in the denomination of our particular chosen num\'eraire.})

Here we consider an attacker who wishes the pools quoted price for the second token, in terms of the first, to deviate from the market prices by a factor of $(1+\epsilon)$.
\emph{Without loss of generality we will consider attacks where the initial price manipulation is done to increase the quoted price of the second token.}
The attacker trades the first token into the pool and withdraws the second tokens.

This is without loss of generality as \emph{a)} we can freely permute the indices of assets in a pool's basket and as \emph{b)} a manipulation to increase the price of one token in the pool, by reducing its reserves, necessarily means we are decreasing the price of the other token.
So an attack based around \emph{decreasing} the price of token, rather than increasing it, is obtained simply by exchanging the indices of the tokens.

Recall Eq~\eqref{eq:tfmm_w_fees}, with fees of $\tau=1-\gamma$ (commonly $\gamma=0.997$), when trading in $\Delta_1>0$ of token $1$ to receive $\Delta_2>0$ of token $2$ the trade must fulfill
\begin{equation}
    (R_1+\gamma\Delta_1)^{w_1}(R_2-\Delta_2)^{w_2}\geq k = R_1^{w_1}R_2^{w_2},
\end{equation}
with the best deal for the trader found when
\begin{equation}
    (R_1+\gamma\Delta_1)^{w_1}(R_2-\Delta_2)^{w_2}= k = R_1^{w_1}R_2^{w_2}.
    \label{eq:fees_trade}
\end{equation}
Eq~\eqref{eq:fees_trade} can be rearranged into a neater form,
\begin{equation}
1-\frac{\D2}{\R2}=\left(1+\gamma\frac{\D1}{\R1}\right)^{-\frac{\w1}{\w2}}.
\label{eq:fees_trade_neat}
\end{equation}

When there are no fees (i.e. $\gamma=1$), the quoted price of token 2 (that is, the ratio of $\Delta_1$ to $\Delta_2$ for an infinitesimal trade of token 1 for token 2) is $m_u:=\frac{R_1}{R_2}\frac{w_2}{w_1}$, and we can expect $m_u=m_p$ at equilibrium.
This is because we can expect arbitrageurs to interact with the pool until $m_u=m_p$.

However, when fees are present this is no longer the case.
In fact, for a given weight vector there is a \emph{range} of reserve values, and thus quoted prices, for which no arbitrage opportunity exists.
We can get the quoted price, for a pool with fees, of the second token with the first token as num\'eraire (again, this is for an infinitesimal trade), by rearranging Eq~\eqref{eq:fees_trade} and taking limits:
\begin{align}
    m_{\mathrm{AMM}}=m_{\mathrm{AMM},2}&=\lim_{\Delta_1,\Delta_2\rightarrow0^+}{\frac{\Delta_1}{\Delta_2}}\nonumber\\
    &=\lim_{\Delta_1\rightarrow0^+} \frac{\Delta_1}{R_2}\left(1-\left(1+\gamma\frac{\Delta_1}{R_1}\right)^{-\frac{w_1}{w_2}}\right)^{-1} \nonumber\\
    \Rightarrow m_{\mathrm{AMM}}=m_{\mathrm{AMM},2}&= \frac{1}{\gamma}\frac{R_1}{R_2}\frac{w_2}{w_1}=\frac{1}{\gamma} m_u.
    \label{eq:prices}
\end{align}
There is \emph{no} arbitrage opportunity when this quoted price is greater than the market price, so when $m_{\mathrm{AMM}}>m_p$.

We also need to consider the flipped trade, where the pool is taking in token 2 and sending out token 1.
As we can simply exchange indices, we get the corresponding price for the first token with the second token as num\'eraire:
\begin{equation}
    m_{\mathrm{AMM},1}= \frac{1}{\gamma}\frac{R_2}{R_1}\frac{w_1}{w_2}.
\end{equation}
Again, there is \emph{no} arbitrage opportunity when $m_{\mathrm{AMM},1}>m_{p,1}=\frac{1}{m_p,2}=\frac{1}{m_p}$.

Putting these together, before the attack we assume that we are at equilibrium---that there is no arbitrage opportunity.
So it must be the case that $\gamma m_p < m_u < \gamma^{-1}m_p$, as $\gamma < 1$ so $\gamma^{-1}>1$.
We are here considering attacks when the attacker is aiming to drive up the quoted price of the second token.
So, the `worst case' then for the pool is for the quoted price for the second token to already be at the extreme upper end of the allowed no-arbitrage range.
We thus have
\begin{align}
    m_u &= \frac{1}{\gamma} m_p\label{eq:mp_mu}\\
    \Rightarrow m_{\mathrm{AMM}}&=\frac{1}{\gamma} m_u= \frac{1}{\gamma^2} m_p
\end{align}
just prior to the attack.

For further mathematical details of the attack see Appendix \ref{app:attack}, where it is shown how both the attacks trades and weight changes have to be `sufficiently large' for attacks to be viable.


\section{Defending Pools}

Our analysis motivates a simple approach to prevent a pool being attackable.
Simply cap trade size, as a proportion of current pool reserves, restrict the maximum change in weight allowed and do not allow any weights to get too small.
Not allowing very large trades can be expected to have minimal to zero impact on the normal operation of TFMM pools, as such trades have large slippage associated with them.
For instance, one could ban trades where reserves of the token being removed would be reduced by 20\%, and ban trades that would increase the reserves of the token being traded in by more than 20\%.
That means we would be choosing $(\bar{\Delta}_1,\bar{\Delta}_2)=(0.2\R1,0.2\R2)$.

Collectively we call these requirements the `\emph{guardrails}':
\begin{itemize}
    \item Restriction on trade size as a proportion of pool reserves
    \item Requirement that weight vector always has entry above a minimum value
    \item Restriction on the absolute values of the entries of the weight change vector
\end{itemize}
Each guardrail has an associated `\emph{guardrail parameter}'.

\subsection{Numerical Simulation}
In this section we describe the results of attempting to perform this attack in simulation.
We are doing this to check that validity of the defence mechanisms we have put forward through studying the problem mathematically in the section above.

We perform a battery of simulated attacks, where the attacker is aiming to maximise their return.
But in this construction, as we are testing whether or not our protection methods work, not only does the attacker get to optimise their trades $\v\Delta$, $\v\Delta'$ but also the state of the pool ($\v\R$, $\v w$), the weight change $\Delta\v w$ and the external market prices $\v m_p$.
We constrain the initial state of the pool and market prices such that no arbitrage opportunity exists before the first trade (i.e. the quoted prices of the pool are within the no-arb region).
The guardrails are applied to the process, and trades have to satisfy the pool's trading function.
We use $\gamma=0.997$ throughout.

Note that naturally we would expect, all else being equal, small maximum trade sizes, large minimum weight values, and small weight changes to be harder to attack.

\begin{figure}[hbtp]
     \centering
     \begin{subfigure}[b]{0.45\columnwidth}
         \centering
         \includegraphics[width=\textwidth]{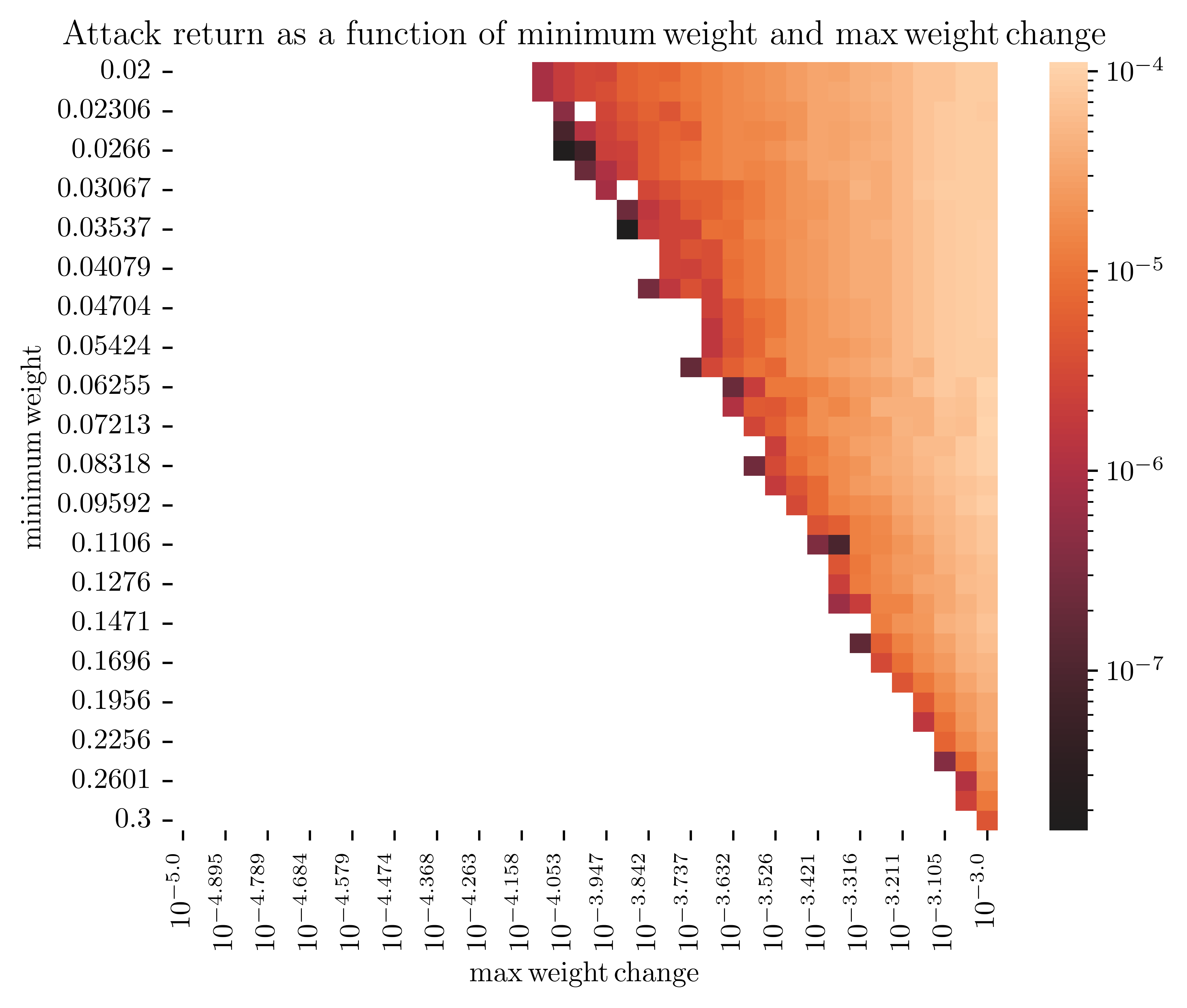}
         \caption{Maximum allowed trade size guardrail fixed at $0.1$ of pool reserves, vary maximum allowed weight change and minimum allowed weight.}
         \label{fig:zoomin}
     \end{subfigure}
     \hfill
     \begin{subfigure}[b]{0.45\columnwidth}
         \centering
         \includegraphics[width=\textwidth]{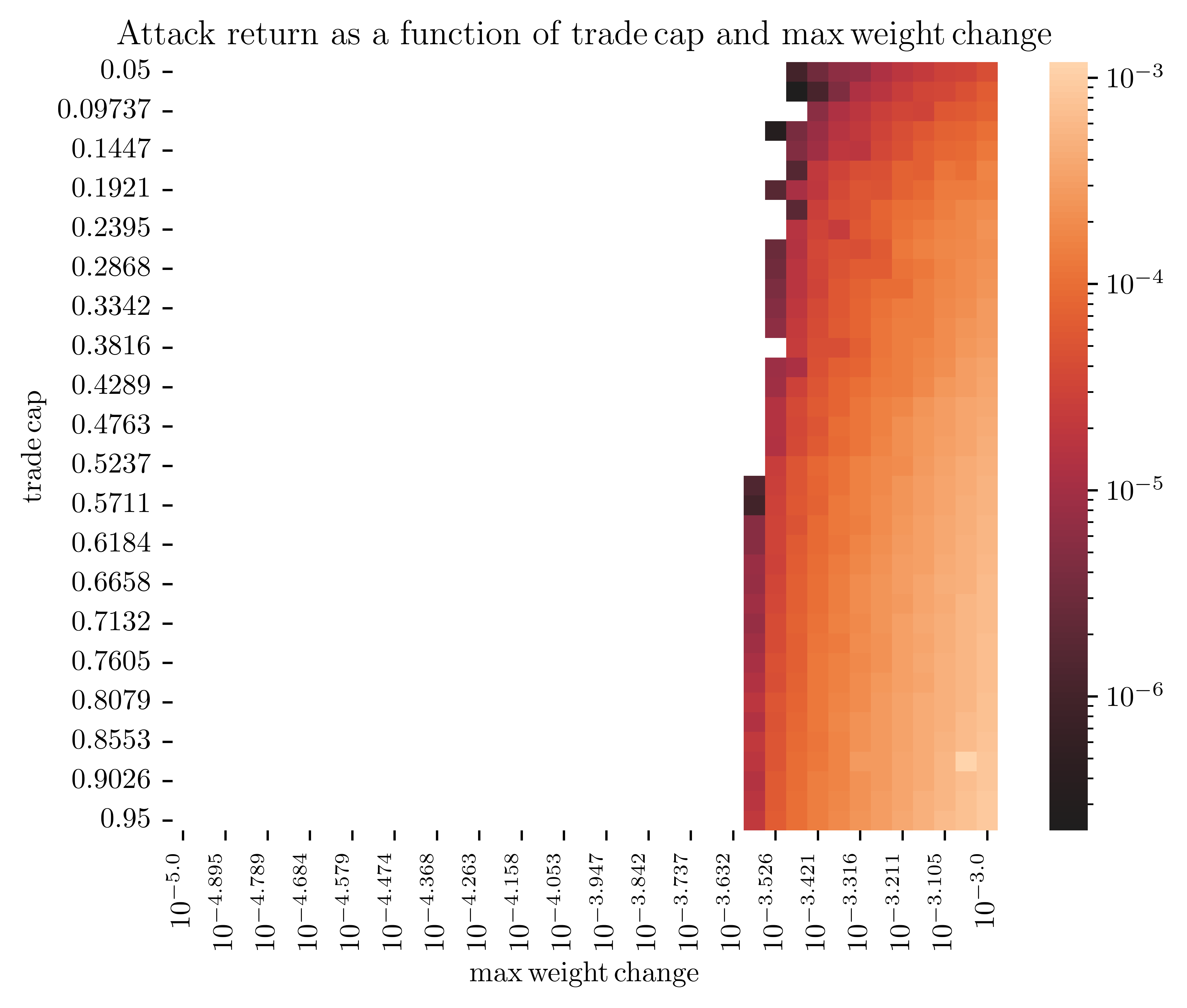}
         \caption{Minimum allowed weight guardrail fixed at $0.1$, vary maximum allowed weight change and maximum allowed trade size.}
         \label{fig:overall}
     \end{subfigure}
     \vspace{1em}
     \centering
     \begin{subfigure}[b]{0.45\columnwidth}
         \centering
         \includegraphics[width=\textwidth]{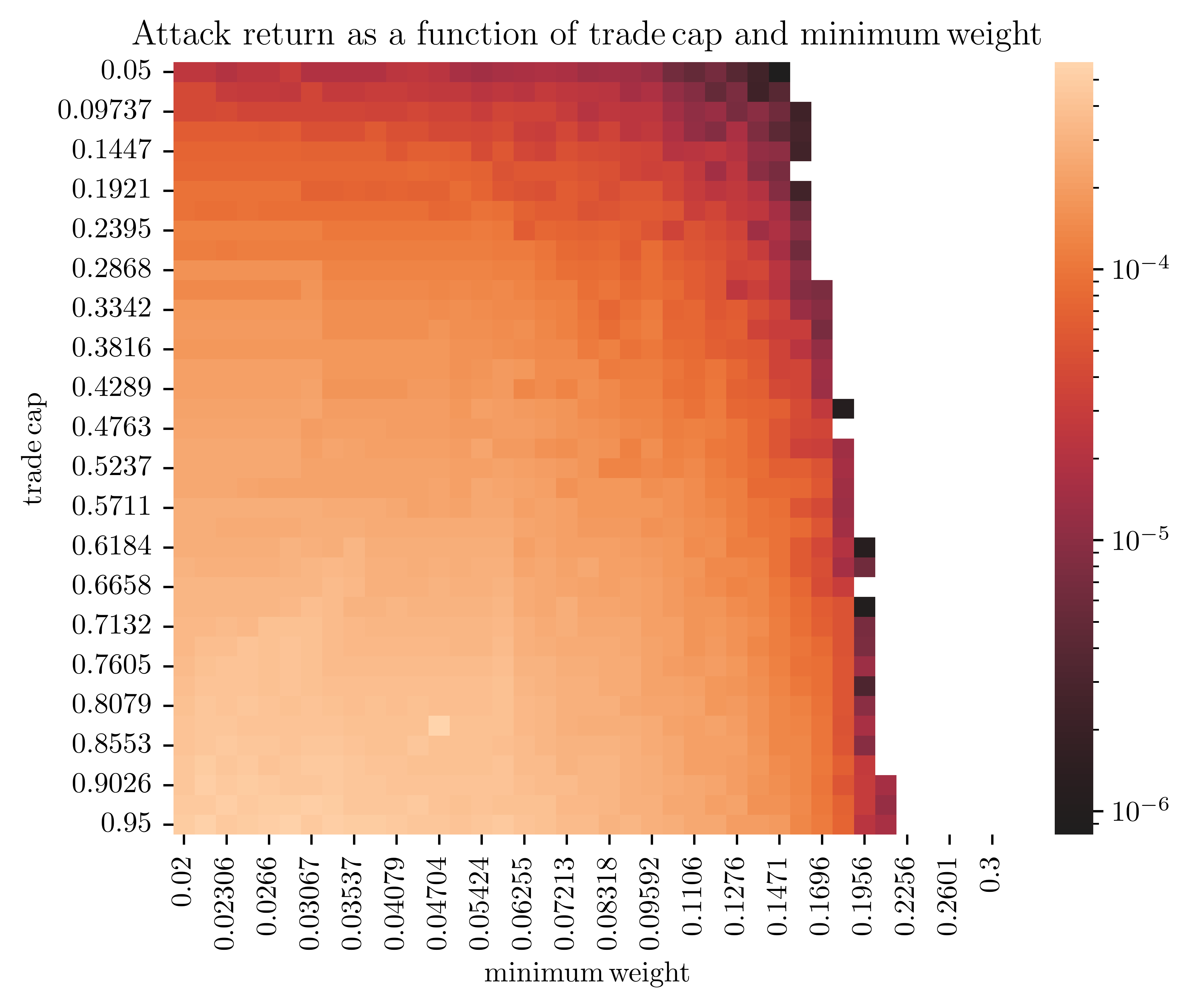}
         \caption{Maximum allowed weight change guardrail fixed at 0.0005, vary minimum allowed weight and maximum allowed trade size.}
         \label{fig:zoomout}
     \end{subfigure}
        \caption{Results from $\approx450,000,000$ attacks, $\approx150,000,000$ per sub-plot, showing the best attack return for each given setting of the guardrails. White indicates that no attack was found, i.e. that the vanilla arbitrage opportunity for the weight change was superior to any found set of potential attack trades. Colour is $Z$ (attack return minus vanilla arbitrage return), scaled by initial pool value.}
        \label{fig:attack}
\end{figure}

\paragraph{Method}
As the problem is non-convex, we use gradient-based optimisation methods to maximise the attacker's return, optimising $\v\Delta$, $\v\R$, $\v w$, $\Delta\v w$ and $\v m_p$.

We make use of the recent advances in methods for finding optimal arbitrage trades in N-token G3M pools~\cite{willetts2024closedform}.
This approach enables for fast, parallelisable and accurate calculation in closed form of the trade that can be applied to a G3M to obtain the most value for the trader, including in the presence of fees, without using a convex solver or other program.
We use these optimal trade results for the attacker's second trade, $\v\Delta'$, the post-weight-change trade, and for the counterfactual arbitrage-only (no manipulation) trade $X(\epsilon_0)$ (see Appendix \ref{app:attack} for more detail).
While $\v\Delta'$ does not require gradient-based optimisation itself, it is key for us that the calculated optimal arbitrage trade is itself auto-differentiable, meaning we can take gradients through it during optimisation.
Trades are not restricted to be one-hot.

We iterate over grids of values of the guardrails.
As three parameters define the guardrails (which are fixed during optimisation), for ease of plotting we do sweeps where one is held constant and the other two vary.
We perform 100,000 separate attacks for each setting of the guardrail parameters, each a slightly different initialisation for the optimisation to act on, and we study 4563 different settings of the guardrail parameters, for $\approx 450,000,000$ different attacks.

We do this for a 3-token pool.
We show the results in Fig~\ref{fig:attack}.
In line with our intuition, pools are harder to attack with larger values of the minimum allowed weight, smaller values of the maximum allowed trade size and weight change.
In this experiment many configuration of guardrail (the white regions in the plots) no attacks were found possible.

\newpage
\section{Conclusion}

This is the first public work studying potential multi-block MEV attacks on dynamic AMMs and defenses thereto, and as such it might have relevance also to Liquidity Bootstrap Pools~\cite{fjordfoundry} and Replicating Market Makers~\cite{angeris2021replicating}.
We used a restrictive method of proof to obtain requirements for robustness to this attack.
With more advanced mathematical methods these bounds could be made tighter, as well as being made to apply to a wider variety of potential set ups, for example analytical results for multi-token attacks (going beyond pair trades).
Additionally there are other potential guardrails and methods for defense beyond those given here.

Further, future refinements to numerical methods may lead to finer understanding of the `attack frontier' (in terms of the guardrail parameters), under which the trade offs between safety and economic performance will only become easier.
\newpage

\bibliography{biblio}
\newpage
\begin{appendices}
\renewcommand\thefigure{\thesection.\arabic{figure}}    
\renewcommand\theequation{\thesection.\arabic{equation}}    

\section{Mathematical Analysis of Potential Attack}
\label{app:attack}

In this potential attack, the attacker first manipulates the pool's quoted price for second token to be
\begin{equation}
    m_{\mathrm{AMM},2}^{\mathrm{manip.}} = (1+\epsilon)m_p,
\end{equation}
where $\epsilon\geq\epsilon_0$.
$\epsilon_0$ is the `do nothing' or `NULL' value---equivalent to their being no attack carried out.
When fees are present and we are in a worst-case scenario for the pool, $\epsilon_0>0$.



$C(\epsilon)$ denotes the cost to the attacker of performing the first, price-manipulating, trade, making clear the dependence of this cost on the scale of the price deviation the attacker does.
$X(\epsilon)$ denotes the return to the attacker from the post-weight-update arbitrage trade.
The overall benefit to the attacker over not carrying out the attack and just being an arbitrageur, $Z(\epsilon)$, is thus
\begin{equation}
    Z(\epsilon)=X(\epsilon) - C(\epsilon) - X(\epsilon_0),
    \label{eq:attack_return}
\end{equation}
When $Z(\epsilon)>0$ the return from the attack, $X(\epsilon) - C(\epsilon)$, is greater than the vanilla, just-doing-arbitrage return $X(\epsilon_0)$.
We can obtain bounds on $X(\epsilon)$ and $C(\epsilon)$ when fees are present without having them in closed form.

\subsection{The stages of the potential attack}
\subsubsection{Cost of Manipulating Quoted Prices}
\label{sec:manip}

Post-trade, the quoted prices are
\begin{equation}
    m_{\mathrm{attacker}} = m_p (1+\epsilon)=  \frac{1}{\gamma}\frac{\frac{\w2}{\R2-\D2}}{\frac{\w1}{\R1+\D1}}.
    \label{eq:attacked_prices}
\end{equation}
Subbing in that $m_p = \gamma m_u$, we find that after the attack trade
\begin{equation}
    \frac{\R1+\D1}{\R2-\D2}=\gamma^2 (1+\epsilon)\frac{\R1}{\R2}.
    \label{eq:post_attack_price}
\end{equation}
Combining Eq~\eqref{eq:post_attack_price} with Eq~\eqref{eq:fees_trade} and rearranging we have that
\begin{equation}
    \left(1+\frac{\D1}{\R1}\right)\left(1+\gamma\frac{\D1}{\R1}\right)^{\frac{\w1}{\w2}} = \gamma^2\left(1+\epsilon\right)
    \label{eq:d1_implicit}
\end{equation}
Similar manipulations give us
\begin{equation}
   \left(1-\frac{\D2}{\R2}\right)^{-1}\left(1+\frac{1}{\gamma}\left(\left(1-\frac{\D2}{\R2}\right)^{-\frac{\w2}{\w1}}-1\right)\right)=\gamma^2\left(1+\epsilon\right).
   \label{eq:d2_implicit}
\end{equation}
$C(\epsilon)$, the \emph{cost} of the attack to the manipulator, again using token 1 as the num\'eraire, is
\begin{equation}
        C(\epsilon) = \D1 - m_p \D2.
        \label{eq:cost}
\end{equation}
Eq~\eqref{eq:d1_implicit} links together $\D1,\R1,
\gamma$ and $\epsilon$, and Eq~\eqref{eq:d2_implicit} separately links $\D2,\R2,
\gamma$ and $\epsilon$.
These are \emph{implicit} equations for $\D1$ or $\D2$ in terms of the other variables.

This means we cannot trivially write down $C(\epsilon)$ in closed form.
We can make progress as we need only either the ratio of $\D1$ to $\D2$ or the \emph{partial derivatives} of the cost, of $\D1$, and of $\D2$, with respect to $\epsilon$ for us to find a bound on $Z(\epsilon)$.

\subsubsection{Change of Reserves of an Attacked Pool After a Weight Update}
\label{sec:weight}
We assume that all this takes place fast enough that the market prices are constant--this could all take place in one block.
After the price manipulation above we have new ($'$ed) reserves
\begin{align}
    \R1'&= \R1 +\D1\, \\
    \R2'&= \R2 - \D2\,
\end{align}
The weights of the pool change, so now we have new (again, $'$ed) weights $\w1'=\w1+\dw1$ and $\w2'=\w2+\dw2$.

After the arbitrage trade, the reserves of the pool will change again (to $''$ed values) such that the new weights, $\w1', \w2'$, and the new post-arbitrage-trade reserves, $\R1'', \R2''$, minimise the value in the pool (thus maximising the returns to the arb).
We thus have
\begin{align}
    \R1''&=\R1'-\D1' \\
    \R2''&=\R2+\D2'.
\end{align}
The return to the arbitrageur is 
\begin{equation}
    X(\epsilon) = \D1' - m_p \D2'
\end{equation}

What is the best-for-the-arb trade?
Instead of directly obtaining the value of $X(\epsilon)$ we will upper bound its value by $X_{\gamma=1}(\epsilon)$ (the return to the arbitrageur when the arbitrageur's trade takes place in a no-fees way---$\gamma=1$ for this trade).
Intuitively $X_{\gamma=1}(\epsilon)>X(\epsilon)$ (after all it would be surprising if fees made the trade cheaper), and in Appendix~\ref{app:gamma_bound} we prove that this is indeed the case.

And thus that
\begin{equation}
m_{\mathrm{AMM}}'=
m_u'=
\frac{\frac{\w2'}{\R2''}}{\frac{\w1'}{\R1''}} =
m_p,
\label{eq:pool_prices_after_arb}
\end{equation}
Combining Eqs~\eqref{eq:post_attack_price} and \eqref{eq:pool_prices_after_arb} we get
\begin{align}
\frac{\frac{\w2'}{\R2''}}{\frac{\w1'}{\R1''}}&=\frac{1}{\gamma}\frac{1}{1+\epsilon}\frac{\frac{\w2}{\R2'}}{\frac{\w1}{\R1'}}\\
\Rightarrow \frac{\R2''}{\R2'}&=\gamma(1+\epsilon)\frac{\w2'}{\w2}\frac{\w1}{\w1'}\frac{\R1''}{\R1'}
\label{eq:star_eq}.
\end{align}



Thus we can consider the no-fees invariant before and after this arb-trade:
\begin{align}
    \tilde{k}' &= \R1'^{\w1'}\R2'^{\w2'}=\R1''^{\w1'}\R2''^{\w2'} \nonumber \\
    \Rightarrow 1&=\left(\frac{\R1''}{\R1'}\right)^{\w1'}\left(\frac{\R2''}{\R2'}\right)^{\w2'}.
\end{align}
Using Eq~\eqref{eq:star_eq} we get
\begin{align}
    1&=\left(\frac{\R1''}{\R1'}\right)^{\w1'}\left(\gamma^2(1+\epsilon)\frac{\w2'}{\w2}\frac{\w1}{\w1'}\frac{\R1''}{\R1'}\right)^{\w2'}, \nonumber \\
    \Rightarrow \frac{\R1''}{\R1'}&=\left(\frac{\w2}{\w2'}\frac{\w1'}{\w1}\frac{1}{\gamma}\frac{1}{1+\epsilon} \right)^{\frac{\w2'}{\w1'+\w2'}},
\end{align}
and thus, similarly, that
\begin{equation}
    \frac{\R2''}{\R2'}=\left(\frac{\w2}{\w2'}\frac{\w1'}{\w1}\frac{1}{\gamma}\frac{1}{1+\epsilon} \right)^{\frac{-\w1'}{\w1'+\w2'}}.
\end{equation}
From algebraic manipulation we obtain that
\begin{align}
    \D1' &= \R1'\left[1-\left(\frac{\w2}{\w2'}\frac{\w1'}{\w1}\frac{1}{\gamma}\frac{1}{1+\epsilon} \right)^{\frac{\w2'}{\w1'+\w2'}}\right],\label{eq:delta1prime}\\
    \D2'&= \R2'\left[\left(\frac{\w2}{\w2'}\frac{\w1'}{\w1}\frac{1}{\gamma}\frac{1}{1+\epsilon} \right)^{\frac{-\w1'}{\w1'+\w2'}}-1\right],\label{eq:delta2prime}
\end{align}
Our upper bound on the return to the arbitrageur is thus
\begin{align}
    X_{\gamma=1}(\epsilon) =& \D1'  - m_p \D2'  \nonumber\\
    \Rightarrow X_{\gamma=1}(\epsilon)=& \R1'\left[1-\left(\frac{\w2}{\w2'}\frac{\w1'}{\w1}\frac{1}{\gamma}\frac{1}{1+\epsilon} \right)^{\frac{\w2'}{\w1'+\w2'}}\right] \nonumber
    \\
    &\quad
    - m_p \R2'\left[\left(\frac{\w2}{\w2'}\frac{\w1'}{\w1}\frac{1}{\gamma}\frac{1}{1+\epsilon} \right)^{\frac{-\w1'}{\w1'+\w2'}}-1\right].
    \label{eq:arb_return}
\end{align}

\subsection{Putting it all together: When is there no extractable value?}

Our upper bound on $Z(\epsilon)$ is thus:
\begin{equation}
    Z(\epsilon)\leq \tilde{Z}(\epsilon) = X_{\gamma=1}(\epsilon) - C(\epsilon) - X_{\gamma=1}(\epsilon_0).
    \label{eq:Zdefn}
\end{equation}


\subsubsection{Bounding via gradients of $Z(\epsilon)$}

Recall that there is a `NULL' value of $\epsilon$, $\epsilon_0$, which corresponds to no-price-manipulation.
As $\D1(\epsilon_0)=\D2(\epsilon_0)=0$, $Z(\epsilon_0)=\tilde{Z}(\epsilon_0)=0$.

We want to find settings of pool parameters such that $Z(\epsilon)<0$ for all $\epsilon>\epsilon_0$.
If $\frac{\partial \tilde{Z}(\epsilon)}{\partial \epsilon}<0$ for all $\epsilon>\epsilon_0$ (i.e. if $Z(\epsilon)$ is a monotonically non-increasing function for $\epsilon>\epsilon_0$) then $Z(\epsilon)<0$ for all $\epsilon>\epsilon_0$.
In the zero fees case the `NULL' value $\epsilon_0=0$.

Taking partial derivatives of $\tilde{Z}(\epsilon)$ w.r.t. $\epsilon$ we get
\begin{equation}
    \frac{\partial \tilde{Z}(\epsilon)}{\partial \epsilon} = \frac{\partial }{\partial \epsilon}\left(\D1'-\D1\right)+ m_p \frac{\partial}{\partial \epsilon}\left(\D2-\D2'\right),
\end{equation}
so if $\frac{\partial }{\partial \epsilon}\left(\D1'-\D1\right)\leq0$ and $\frac{\partial}{\partial \epsilon}\left(\D2-\D2'\right)\leq0$, then we can guarantee that the attack will not work.


\paragraph{Gradient of ${\D1'-\D1}$ w.r.t. $\epsilon$}

Using Eq~\eqref{eq:delta1prime}, recalling that $\R1'=\R1+\D1$, we have that
\begin{align}
    \D1'-\D1 &= \R1 - \left(\frac{\w2}{\w2'}\frac{\w1'}{\w1}\frac{1}{\gamma}\frac{1}{1+\epsilon} \right)^{\frac{\w2'}{\w1'+\w2'}}\left(\D1+\R1\right)\\
    \Rightarrow \frac{\partial }{\partial \epsilon}\left(\D1'-\D1\right) &= \left(\frac{\w2}{\w2'}\frac{\w1'}{\w1}\frac{1}{\gamma}\frac{1}{1+\epsilon} \right)^{\frac{\w2'}{\w1'+\w2'}}\nonumber \\
    &\quad \times \left(\frac{1}{1+\epsilon}{\frac{\w2'}{\w1'+\w2'}}\left(\D1+\R1\right)-\frac{\partial\D1}{\partial\epsilon}\right)\label{eq:1deriv}.
\end{align}
We find that (see Appendix \ref{app:d1})
\begin{equation}
    \frac{\partial\D1}{\partial\epsilon} = \frac{\gamma^2\R1}{\left(1+\gamma\frac{\w1}{\w2}\left(1+\frac{\D1}{\R1}\right)\left(1+\gamma\frac{\D1}{\R1}\right)^{-1}\right)\left(1+\gamma\frac{\D1}{\R1}\right)^{\frac{\w1}{\w2}}}.
\end{equation}
Subbing this into Eq~\eqref{eq:1deriv}, and using Eq~\eqref{eq:d1_implicit}, for $\frac{\partial }{\partial \epsilon}\left(\D1'-\D1\right)\leq0$ it must be that
\begin{equation}
    {\frac{\w2'}{\w1'+\w2'}}\left(1+\gamma\frac{\w1}{\w2}\left(1+\frac{\D1}{\R1}\right)\left(1+\gamma\frac{\D1}{\R1}\right)^{-1}\right)\leq1,\label{eq:w2restrict}
\end{equation}
where we have used that $\left(\frac{\w2}{\w2'}\frac{\w1'}{\w1}\frac{1}{\gamma}\frac{1}{1+\epsilon} \right)^{\w2'}>0$ when $\w1\in(0,1)$, $\w2\in(0,1)$, $\w1'\in(0,1)$, $\w2'\in(0,1)$, $0<\gamma<1$, and $\epsilon>\epsilon_0$.

\paragraph{Gradient of ${\D2-\D2'}$ w.r.t. $\epsilon$}
Using Eq~\eqref{eq:delta2prime}, recalling that $\R2'=\R2-\D2$, we have that
\begin{align}
    \D2-\D2' &= \R2 + \left(\frac{\w2}{\w2'}\frac{\w1'}{\w1}\frac{1}{\gamma}\frac{1}{1+\epsilon} \right)^{-\w1'}\left(\D2-\R2\right)\\
    \Rightarrow \frac{\partial }{\partial \epsilon}\left(\D2-\D2'\right) &= \left(\frac{\w2}{\w2'}\frac{\w1'}{\w1}\frac{1}{\gamma}\frac{1}{1+\epsilon} \right)^{\frac{-\w1'}{\w1'+\w2'}}\nonumber\\
    &\quad\times\left(\frac{1}{1+\epsilon}{\frac{\w1'}{\w1'+\w2'}}\left(\D2-\R2\right)+\frac{\partial\D2}{\partial\epsilon}\right)\label{eq:2deriv}.
\end{align}
We find that (see Appendix \ref{app:d2})
\begin{equation}
    \frac{\partial\D2}{\partial\epsilon} = \frac{\gamma^3\R2\left(1-\frac{\D2}{\R2}\right)^2}{\left(1+\frac{\w2}{\w1}\right)\left(1-\frac{\D2}{\R2}\right)^{-\frac{\w2}{\w1}}-\left(1-\gamma\right)}
\end{equation}
Subbing this into Eq~\eqref{eq:2deriv}, and using Eq~\eqref{eq:d2_implicit}, for $\frac{\partial }{\partial \epsilon}\left(\D2-\D2'\right)\leq0$ it must be that
\begin{equation}
    {\frac{\w1'}{\w1'+\w2'}}\geq\frac{1-\left(1-\gamma\right)\left(1-\frac{\D2}{\R2}\right)^{\frac{\w2}{\w1}}}{1+\frac{\w2}{\w1}-\left(1-\gamma\right)\left(1-\frac{\D2}{\R2}\right)^{\frac{\w2}{\w1}}}\label{eq:w1restrict},
\end{equation}
where we have used that $\left(\frac{\w2}{\w2'}\frac{\w1'}{\w1}\frac{1}{\gamma}\frac{1}{1+\epsilon} \right)^{-\w1'}>0$ when $\w1\in(0,1)$, $\w2\in(0,1)$, $\w1'\in(0,1)$, $\w2'\in(0,1)$, $0<\gamma<1$, and $\epsilon>\epsilon_0$.

These results tell us that if the changes in weights are within these bounds, then no attack is possible.

\paragraph{2-token case}

For the case of a two-token pool, so the two tokens being traded are the old tokens present, we can simplify the above equations and plot them.
We are interested in knowing what weight changes we can `get away with' without a pool being open to attack.
That means we are most interested in the above inequalities reformulated explicitly to give us bounds on the weight changes.
As we are now in the two-token case, $\w1'+\w2'=\w1+\w2=1$ and $\Delta \w1 = -\Delta \w2$.
Thus we can re-write Eq~\eqref{eq:w2restrict} and Eq~\eqref{eq:w1restrict} in terms of just $w = \w1=1-\w2$ and $\Delta w = \Delta\w1=-\Delta\w2$ :
\begin{align}
    \Delta w &\geq1- \frac{1}{\left(1+\gamma\frac{w}{\left(1-w\right)}\left(1+\frac{\D1}{\R1}\right)\left(1+\gamma\frac{\D1}{\R1}\right)^{-1}\right)}-w,\label{eq:w2restrict_rewrite}\\
\Delta w &\geq\frac{1-\left(1-\gamma\right)\left(1-\frac{\D2}{\R2}\right)^{\frac{\left(1-w\right)}{w}}}{1+\frac{\left(1-w\right)}{w}-\left(1-\gamma\right)\left(1-\frac{\D2}{\R2}\right)^{\frac{\left(1-w\right)}{w}}}-w\label{eq:w1restrict_rewrite}.
\end{align}
In \S\ref{app:attack} we described how we could study the case where the attacker initially increases the quoted price of the second token, denominated in the first, \emph{without loss of generality} as we can simply swap indices in our final equations to get the results for the mirror-attack where the price of the first token is initially pumped up by the attacker.
We can now do that swapping, giving the additional constraints:
\begin{align}
    \Delta w &\leq \frac{1}{\left(1+\gamma\frac{\left(1-w\right)}{w}\left(1+\frac{\D2}{\R2}\right)\left(1+\gamma\frac{\D2}{\R2}\right)^{-1}\right)}-w,\label{eq:w2restrict_reindex}\\
\Delta w &\leq1-\frac{1-\left(1-\gamma\right)\left(1-\frac{\D1}{\R1}\right)^{\frac{w}{\left(1-w\right)}}}{1+\frac{w}{\left(1-w\right)}-\left(1-\gamma\right)\left(1-\frac{\D1}{\R1}\right)^{\frac{w}{\left(1-w\right)}}}-w\label{eq:w1restrict_reindex}.
\end{align}

If the changes in weight fulfil Eqs~(\ref{eq:w2restrict_rewrite}-\ref{eq:w1restrict_reindex}) then no attack is possible for a given initial price-deviating trade.
These bounds are expressed as functions of that initial trade, $\D1$ and $\D2$.
These bounds are also monotonic in $\D1$ and $\D2$ such that if a pool is safe under an initial trade $(\bar{\Delta}_1,\bar{\Delta}_2)$ then it is safe also for all trades $\D1<\bar{\Delta}_1$ and $\D2<\bar{\Delta}_2$.
See Appendix \ref{app:monotone} for this.

\begin{figure}
    \centering
    \includegraphics[width=0.7\columnwidth]{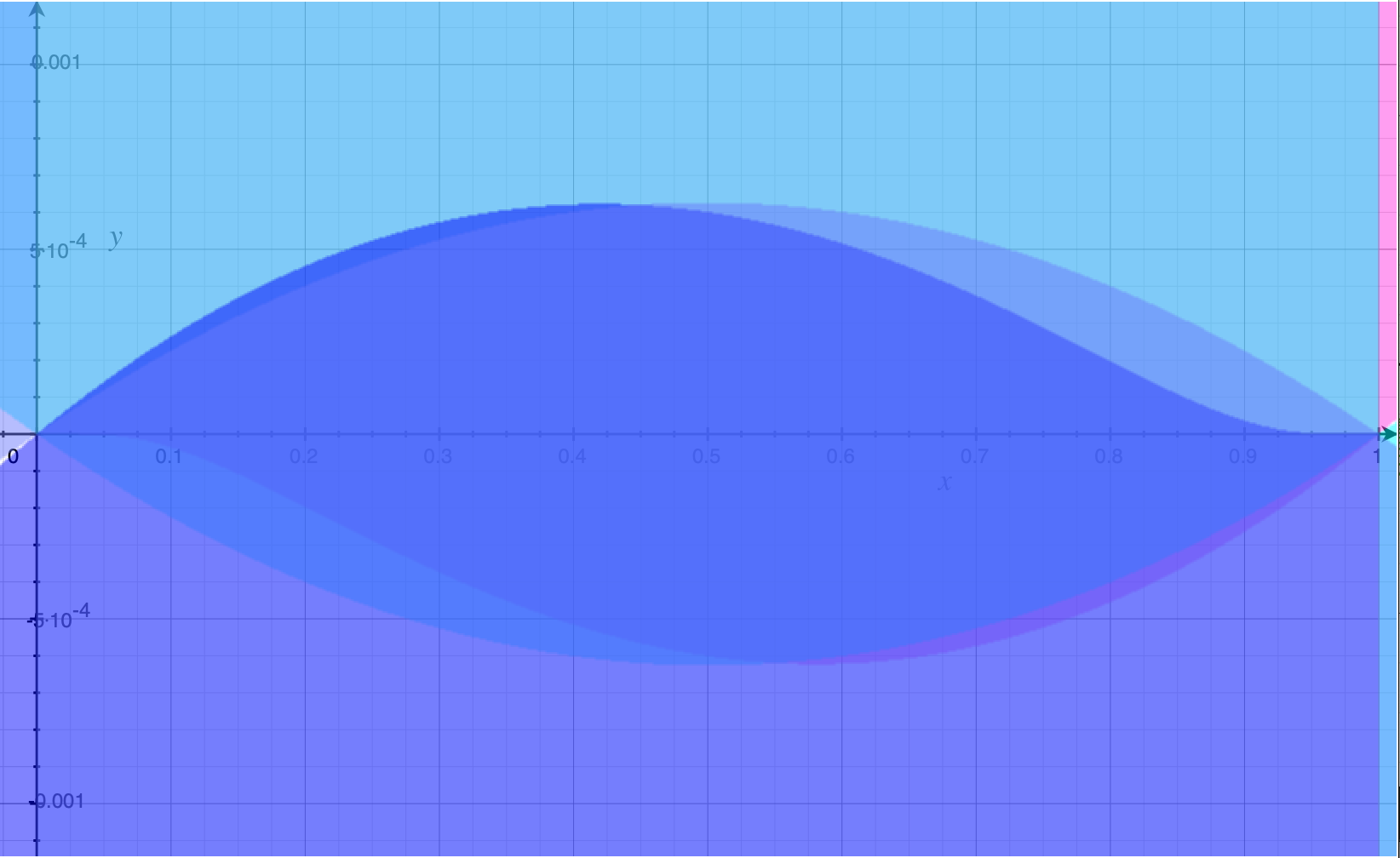}
    \caption{Plots of the two-token inequalities Eqs~(\ref{eq:w2restrict_rewrite}-\ref{eq:w1restrict_reindex}), where shaded regions are the safe region of each inequality.
    This is done for $\Delta_1=0.2 \R1$ and $\gamma=0.997$. The x-axis is $w$ and the y-axis is $\Delta w$.}
    \label{fig:numerical_bounds}
\end{figure}

\subsection{Proof that $X_{\gamma=1}(\epsilon)$ is an upper bound on $X(\epsilon)$}
\label{app:gamma_bound}
Here we will demonstrate that $X_{\gamma=1}(\epsilon) > X(\epsilon)$ for $\epsilon>\epsilon_0$.
First recall that
\begin{equation}
    X(\epsilon) = \D1'(\epsilon,\gamma) - m_p \D2'(\epsilon,\gamma)
    \nonumber
\end{equation}
and that
\begin{equation}
    X_{\gamma=1}(\epsilon) = \D1'(\epsilon,\gamma=1) - m_p \D2'(\epsilon,\gamma=1),
    \nonumber
\end{equation}
where we have made explicit the dependence of $\D1'$ and $\D2'$ on $\epsilon$ and $\gamma$.
\begin{equation}
    X_{\gamma=1}(\epsilon) - X(\epsilon) = \D1'(\epsilon,\gamma=1) - \D1'(\epsilon,\gamma) - \left(m_p \D2'(\epsilon,\gamma=1)-\D2'(\epsilon,\gamma)\right),
    \nonumber
\end{equation}
This means for $X_{\gamma=1}(\epsilon)>X(\epsilon)$ it is sufficient to show that $\D2'(\epsilon,\gamma)>\D2'(\epsilon,\gamma=1)$ and that $\D1'(\epsilon,\gamma)<\D1'(\epsilon,\gamma=1)$.
We will handle these in turn

\paragraph{Showing $\D2'(\epsilon,\gamma)>\D2'(\epsilon,\gamma=1)$:}

We begin by writing down the trade-invariant for $\D1'$ and $\D2'$ in the presence of fees.
It is, naturally, that
\begin{equation}
    (\R1'-\D1'(\epsilon,\gamma))^{\w1'}(\R2'+\gamma\D2'(\epsilon,\gamma))^{\w2'}= k'=\R1'^{\w1'}\R2'^{\w2},
    \label{eq:ap_fees_trade}
\end{equation}
as the trader is putting $\D2'$ of token 2 into the pool and withdrawing $\D1'$ of token 1.

Next, we need the trade invariant if $\gamma=1$---if there are no fees.
Then
\begin{equation}
    (\R1'-\D1'(\epsilon,\gamma=1))^{\w1'}(\R2'+\D2'(\epsilon,\gamma=1))^{\w2'}= k'=\R1'^{\w1'}\R2'^{\w2}.
    \label{eq:ap_nofees_trade}
\end{equation}

The final part we need is the following.
We know that the purpose of this trade is to get the pool to quote a certain price after the trade has been performed, based on its post-trade pool reserves.
This means that the quoted prices after the trade-with-fees will, indeed must, be the same as the quoted price after the trade-with-no-fees.\footnote{Note that we are only setting $\gamma=1$ \emph{for the trade} in the $\gamma=1$ part of this construction.
We are still having the arbitrage trade, with fees or not, bring the quoted price to the upper end of the ($\gamma$ defined) no-arb region.}
Thus, using Eq~\eqref{eq:pool_prices_after_arb} we get that
\begin{equation}
    \frac{(\R1'-\D1'(\epsilon,\gamma=1))}{(\R2'+\D2'(\epsilon,\gamma=1))}=\frac{(\R1'-\D1'(\epsilon,\gamma))}{(\R2'+\D2'(\epsilon,\gamma))},
\label{eq:app_price_equality}
\end{equation}
where the weights have cancelled out.
Rearranging we get that
\begin{equation}
    {(\R1'-\D1'(\epsilon,\gamma))}={(\R2'+\D2'(\epsilon,\gamma))}\frac{(\R1'-\D1'(\epsilon,\gamma=1))}{(\R2'+\D2'(\epsilon,\gamma=1))},
\nonumber
\end{equation}
which we can sub in to Eq~\eqref{eq:ap_fees_trade} to get
\begin{equation}
    \left({(\R2'+\D2'(\epsilon,\gamma))}\frac{(\R1'-\D1'(\epsilon,\gamma=1))}{(\R2'+\D2'(\epsilon,\gamma=1))}\right)^{\w1'}(\R2'+\gamma\D2'(\epsilon,\gamma))^{\w2'}= k'.
    \nonumber
\end{equation}
Rearranging Eq~\eqref{eq:ap_nofees_trade} we get
\begin{equation}
    (\R1'-\D1'(\epsilon,\gamma=1))^{\w1'}= \frac{k'}{(\R2'+\D2'(\epsilon,\gamma=1))^{\w2'}},
    \nonumber
\end{equation}
which we can then sub in to the previous equation to obtain
\begin{equation}
    {{(\R2'+\D2'(\epsilon,\gamma))^{\w1'}}(\R2'+\gamma\D2'(\epsilon,\gamma))^{\w2'}}={(\R2'+\D2'(\epsilon,\gamma=1))}.
    \nonumber
\end{equation}
As $\w1'+\w2'=1$, $0<\w1'<1$, $0<\w2'<1$ and $0<\gamma<1$, it is thus clear that $\D2'(\epsilon,\gamma)>\D2'(\epsilon,\gamma=1)$.

\paragraph{Showing $\D1'(\epsilon,\gamma)<\D1'(\epsilon,\gamma=1)$:}
Rearranging Eq~\eqref{eq:app_price_equality} we get
\begin{align}
    \D1'(\epsilon,\gamma=1)&=\R1'-{(\R1'-\D1'(\epsilon,\gamma))}\frac{(\R2'+\D2'(\epsilon,\gamma=1))}{(\R2'+\D2'(\epsilon,\gamma))},\nonumber\\
\Rightarrow  \D2'(\epsilon,\gamma=1) &=\R1'(1-k)+k\D1'(\epsilon,\gamma),\nonumber\\
\Rightarrow  \D2'(\epsilon,\gamma) &=\frac{1}{k}\D1'(\epsilon,\gamma=1)+\R1'\frac{(k-1)}{k},\nonumber
\end{align}
where $k=\frac{(\R2'+\D2'(\epsilon,\gamma=1))}{(\R2'+\D2'(\epsilon,\gamma))}$.
$k<1$ as $\D2'(\epsilon,\gamma)>\D2'(\epsilon,\gamma=1)$ and $k>0$ as both its numerator and denominator are $>0$.
This last equation, for $\D1'(\epsilon,\gamma)$ as a function of $\D1'(\epsilon,\gamma=1)$, is a straight line with gradient $1/k>1$ and intercept $\R1'\frac{(k-1)}{k}<0$.
This line crosses `$y=x$' when $\R1'=\D1'(\epsilon,\gamma=1)$, so for all $\D1'(\epsilon,\gamma=1)<\R1'$ (which are the only possible values as the pool cannot be drained) we have that $\D1(\epsilon,\gamma=1)<\D1'(\epsilon,\gamma=1)$ as required.
\subsection{Finding partial derivatives}
\subsubsection{Finding $\partial \D1 / \partial \epsilon$}
\label{app:d1}
Recall the implicit equation that defines $\D1$, Eq~\eqref{eq:d1_implicit}:
\begin{equation}
    \left(1+\frac{\D1}{\R1}\right)\left(1+\gamma\frac{\D1}{\R1}\right)^{\frac{\w1}{\w2}} = \gamma^2\left(1+\epsilon\right).
    \nonumber
\end{equation}
Taking partial derivatives of both sides with respect to $\epsilon$:
\begin{align}
    &\frac{\partial}{\partial\epsilon}\left(\left(1+\frac{\D1}{\R1}\right)\left(1+\gamma\frac{\D1}{\R1}\right)^{\frac{\w1}{\w2}}\right)= \frac{\partial}{\partial\epsilon}\left(\gamma^2\left(1+\epsilon\right)\right)\nonumber\\
    \Rightarrow &\frac{1}{\R1}\frac{\partial\D1}{\partial\epsilon}\left(1+\gamma\frac{\D1}{\R1}\right)^{\frac{\w1}{\w2}}+\left(1+\frac{\D1}{\R1}\right)\frac{\w1}{\w2}\left(1+\gamma\frac{\D1}{\R1}\right)^{\frac{\w1}{\w2}-1}\frac{\gamma}{\R1}\frac{\partial\D1}{\partial\epsilon}=\gamma^2\nonumber\\
    \Rightarrow&\frac{\partial\D1}{\partial\epsilon}\frac{1}{\R1}\left(1+\gamma\frac{\D1}{\R1}\right)^{\frac{\w1}{\w2}}\left(1+\gamma\left(1+\frac{\D1}{\R1}\right)\frac{\w1}{\w2}\left(1+\gamma\frac{\D1}{\R1}\right)^{-1}\right)=\gamma^2\nonumber\\
    \Rightarrow&\frac{\partial\D1}{\partial\epsilon}=\frac{\gamma^2\R1}{\left(1+\gamma\frac{\w1}{\w2}\left(1+\frac{\D1}{\R1}\right)\left(1+\gamma\frac{\D1}{\R1}\right)^{-1}\right)\left(1+\gamma\frac{\D1}{\R1}\right)^{\frac{\w1}{\w2}}}\nonumber,
\end{align}
as required.
\subsubsection{Finding $\partial \D2 / \partial \epsilon$}
\label{app:d2}
Recall the implicit equation that defines $\D2$, Eq~\eqref{eq:d2_implicit}:
\begin{equation}
   \left(1-\frac{\D2}{\R2}\right)^{-1}\left(1+\frac{1}{\gamma}\left(\left(1-\frac{\D2}{\R2}\right)^{-\frac{\w2}{\w1}}-1\right)\right)=\gamma^2\left(1+\epsilon\right).
   \nonumber
\end{equation}
Taking partial derivatives of both sides with respect to $\epsilon$:
\begin{align}
    &\frac{\partial}{\partial\epsilon}\left(\left(1-\frac{\D2}{\R2}\right)^{-1}\left(1+\frac{1}{\gamma}\left(\left(1-\frac{\D2}{\R2}\right)^{-\frac{\w2}{\w1}}-1\right)\right)\right)=\frac{\partial}{\partial\epsilon}\left(\gamma^2\left(1+\epsilon\right)\right)
   \nonumber\\
   \Rightarrow&\frac{\partial\D2}{\partial\epsilon}\frac{1}{\R2}\left(1-\frac{\D2}{\R2}\right)^{-2}\left(\left(1+\frac{1}{\gamma}\Bigg(1-\frac{\D2}{\R2}\right)^{-\frac{\w2}{\w1}}-\frac{1}{\gamma}\right)\nonumber\\
   &\quad\quad+\left(\frac{1}{\gamma}\frac{\w2}{\w1}\left(1-\frac{\D2}{\R2}\right)^{-\frac{\w2}{\w1}}\right)\Bigg)=\gamma^2\nonumber\\
   \Rightarrow&\frac{\partial\D2}{\partial\epsilon}=\frac{\gamma^3\R2\left(1-\frac{\D2}{\R2}\right)^2}{\left(1+\frac{\w2}{\w1}\right)\left(1-\frac{\D2}{\R2}\right)^{-\frac{\w2}{\w1}}-\left(1-\gamma\right)}\nonumber,
\end{align}
as required.
\subsection{Monotonicity of inequalities}
\label{app:monotone}
We have two inequalities that we derive above, Eq~\eqref{eq:w2restrict},
\begin{equation}
    {\frac{\w2'}{\w1'+\w2'}}\left(1+\gamma\frac{\w1}{\w2}\left(1+\frac{\D1}{\R1}\right)\left(1+\gamma\frac{\D1}{\R1}\right)^{-1}\right)\leq1,\nonumber
\end{equation}
and Eq~\eqref{eq:w1restrict},
\begin{equation}
    {\frac{\w1'}{\w1'+\w2'}}\geq\frac{1-\left(1-\gamma\right)\left(1-\frac{\D2}{\R2}\right)^{\frac{\w2}{\w1}}}{1+\frac{\w2}{\w1}-\left(1-\gamma\right)\left(1-\frac{\D2}{\R2}\right)^{\frac{\w2}{\w1}}}\nonumber.
\end{equation}
Here we are interested in the monotonicity of the level-sets of these w.r.t. $\D1$ and $\D2$.
That is, if our values of $\w1',\w2'$ satisfy their inequalities for some particular values of $\w1,\w2,\R1,\R2,\D1,\D2$, can we guarantee that the same values of $\w1',\w2'$ also satisfy the inequalities for $\hat{\Delta}_1<\D1$ and $\hat{\Delta}_2<\D2$ (with $\w1,\w2,\R1,\R2$ fixed)?

For this to be the case, we need the level sets of the inequalities to always have the correct gradient so that smaller values of $\D1,\D2$ always move the inequality's boundary `further away' from the value of $\w1',\w2'$.
Let us handle the above equations in turn.
\paragraph{$\D1$}

Consider the case where our $\w1',\w2'$ values just satisfies the first inequality above, so that
\begin{equation}
    {\frac{\bar{w}_2'}{\bar{w}_1'+\bar{w}_2'}}=\frac{1}{1+\gamma\frac{\w1}{\w2}\left(1+\frac{\D1}{\R1}\right)\left(1+\gamma\frac{\D1}{\R1}\right)^{-1}},\nonumber,
\end{equation}
where $\bar{w}_1', \bar{w}_2'$ denotes these critical value of $\w1',\w2'$.
For $\bar{w}_1', \bar{w}_2'$ to also satisfy the inequality for $\hat{\Delta}_1<\D1$, we need that
\[\frac{\partial}{\partial\D1}\left(\frac{1}{1+\gamma\frac{\w1}{\w2}\left(1+\frac{\D1}{\R1}\right)\left(1+\gamma\frac{\D1}{\R1}\right)^{-1}}\right)<0,\]
as then the required critical value will be larger for smaller $\D1$ values.
Evaluating this partial derivative, we find that
\begin{align}
    \frac{\partial}{\partial\D1}&\left(\frac{1}{1+\gamma\frac{\w1}{\w2}\left(1+\frac{\D1}{\R1}\right)\left(1+\gamma\frac{\D1}{\R1}\right)^{-1}}\right)\nonumber\\
    &\quad=-\frac{\frac{\gamma\w1}{\w2}\left(1+\gamma\frac{\D1}{\R1}\right)^{-1}\left(1+\gamma\left(1+\frac{\D1}{\R1}\right)\left(1+\gamma\frac{\D1}{\R1}\right)^{-1}\right)}{\left(1+\gamma\frac{\w1}{\w2}\left(1+\frac{\D1}{\R1}\right)\left(1+\gamma\frac{\D1}{\R1}\right)^{-1}\right)^2}.
\end{align}
For $\w1\in(0,1)$, $\w2\in(0,1)$, $\gamma>0$, $\R1>0$ and $\D1>0$, clearly this gradient is always negative, as required.

\paragraph{$\D2$}
Again let us take the critical value of the new weight such that the inequality is just satisfied, so
\begin{equation}
    \frac{\bar{w}_1'}{\bar{w}_1'+\bar{w}_2'}=\frac{1-\left(1-\gamma\right)\left(1-\frac{\D2}{\R2}\right)^{\frac{\w2}{\w1}}}{1+\frac{\w2}{\w1}-\left(1-\gamma\right)\left(1-\frac{\D2}{\R2}\right)^{\frac{\w2}{\w1}}}\label{eq:w1_critical}.
\end{equation}
Here we want the partial derivative of this critical value to always be positive, so that the required critical value is always smaller for smaller $\D2$ values, so we want that
\[\frac{\partial}{\partial\D2}\left(\frac{1-\left(1-\gamma\right)\left(1-\frac{\D2}{\R2}\right)^{\frac{\w2}{\w1}}}{1+\frac{\w2}{\w1}-\left(1-\gamma\right)\left(1-\frac{\D2}{\R2}\right)^{\frac{\w2}{\w1}}}\right)>0.\]
We can re-write Eq~\eqref{eq:w1_critical} more compactly as
\begin{equation}
    \frac{\bar{w}_1'}{\bar{w}_1'+\bar{w}_2'}=\frac{1-f(\D2)}{1+a-f(\D2)}\label{eq:w1_critical_rewrite}.
\end{equation}
where $f(\D2)=\left(1-\gamma\right)\left(1-\frac{\D2}{\R2}\right)^{\frac{\w2}{\w1}}$ and $a=\frac{\w2}{\w1}$.
\[\frac{\partial}{\partial\D2}\left(\frac{1-f(\D2)}{1+a-f(\D2)}\right)=-\frac{a\frac{\partial f(\D2)}{\partial\D2}}{\left(1+a-f(\D2)\right)^2}.\]
As $a>0$, as $\w1\in(0,1)$ and $\w2\in(0,1)$, the gradient $\frac{\partial\bar{w}_1'}{\partial\D2}$ is positive if $\frac{\partial f(\D2)}{\partial\D2}<0$.
Let us evaluate $\frac{\partial f(\D2)}{\partial\D2}$:
\begin{align}
\frac{\partial f(\D2)}{\partial \D2}&=\frac{\partial }{\partial \D2}\left(\left(1-\gamma\right)\left(1-\frac{\D2}{\R2}\right)^{\frac{\w2}{\w1}}\right) \nonumber \\
&=\left(-\frac{1}{\R2}\frac{\w2}{\w1}\left(1-\gamma\right)\left(1-\frac{\D2}{\R2}\right)^{\frac{\w2}{\w1}-1}\right) \nonumber
\end{align}
As $\R2<\D2$, $\w1\in(0,1)$, $\w2\in(0,1)$, $0<\gamma<1$, this gradient is always negative, so\\
$\frac{\partial}{\partial\D2}\left(\frac{1-\left(1-\gamma\right)\left(1-\frac{\D2}{\R2}\right)^{\frac{\w2}{\w1}}}{1+\frac{\w2}{\w1}-\left(1-\gamma\right)\left(1-\frac{\D2}{\R2}\right)^{\frac{\w2}{\w1}}}\right)>0$, as required.

\end{appendices}
\newpage
\textbf{DISCLAIMER} This paper is for general information purposes only.
It does not constitute investment advice or a recommendation or solicitation to buy or sell any investment or asset, or participate in systems that use TFMM; nor is it a guarantee in any form of the behaviour or performance of any TFMM-based systems.
This paper should not be used in the evaluation of the merits of making any investment decision.
It should not be relied upon for accounting, security, legal or tax advice or investment recommendations.
This paper reflects current opinions of the authors regarding the development and functionality of TFMM and is subject to change without notice or update.

While some aspects, such as altering target weights in geometric mean market makers is prior art, aspects of TFMM that are novel for use in dynamic weight AMMs, for purposes of core liquidity providing or forms of asset management including, but not exclusively, fund construction, structured products, treasury management are covered by patent filing date of 21st February 2023.
\end{document}